\documentclass[twocolumn,aps,floats]{revtex4}
\usepackage{amsmath}
\usepackage{graphicx}
\usepackage{float}    

\newcommand{\mr}[1]{{{\mathrm{#1}}}}
\newcommand{\mcal}[1]{{\mathcal{#1}}}

\newcommand{\dt}{\partial_\tau}
\newcommand{\inte}{\int_0^\beta \!\!\!\! \mr{d}\tau}

\newcommand{\w}{\omega}
\newcommand{\s}{\sigma}
\newcommand{\x}{{\bf x}}
\newcommand{\0}{{\bf 0}}
\newcommand{\1}{{\bf 1}}
\newcommand{\cs}{c^{\phantom{\dagger}}_\s}
\newcommand{\csp}{c^{\phantom{\dagger}}_{\s'}}
\newcommand{\csd}{c^{\dagger}_\s}
\newcommand{\csb}{{\bar c}^{\phantom\dagger}_\s}

\newcommand{\bs}{b^{\phantom{\dagger}}_\s}

\newcommand{\bsd}{b^{\dagger}_\s}
\newcommand{\bsb}{{\bar b}^{\phantom\dagger}_\s}

\newcommand{\fm}{f^{\phantom{\dagger}}_m}

\newcommand{\fmd}{f^{\dagger}_m}
\newcommand{\fmb}{{\bar f}^{\phantom\dagger}_m}

\newcommand{\cks}{c^{\phantom{\dagger}}_{k\s}}
\newcommand{\cksd}{c^{\dagger}_{k\s}}
\newcommand{\cksb}{{\bar c}^{\phantom\dagger}_{k\s}}

\newcommand{\e}{\epsilon}
\newcommand{\ek}{\epsilon_k}
\newcommand{\Simp}{S_\mr{imp}}
\newcommand{\Cimp}{\chi_\mr{imp}}
\newcommand{\JB}{J_B}

\begin{document}

\title{Critical points and non-Fermi liquids in the underscreened pseudogap Kondo model}

\author{Serge Florens and Matthias Vojta}
\affiliation{\mbox{Institut f\"ur Theorie der Kondensierten
Materie, Universit\"at Karlsruhe, 76128 Karlsruhe, Germany}}

\date{\today}

\begin{abstract}
Numerical Renormalization Group simulations have shown that
the underscreened spin-1 Kondo impurity model with
power-law bath density of states (DOS) $\rho(\w) \propto |\w|^r$
possesses various intermediate-coupling fixed points,
including a stable non-Fermi liquid phase.
In this paper we discuss the corresponding universal low-energy theories,
obtain thermodynamic quantities and critical exponents
by renormalization group analysis together with suitable
$\epsilon$-expansions, and compare our results with numerical
data.
Whereas the particle-hole symmetric critical point can be controlled at weak coupling
using a simple generalization of the spin-1/2 model,
we show that the stable non-Fermi liquid fixed point must be accessed
near strong coupling via a mapping onto an effective ferromagnetic
$S_\mr{eff}=1/2$ model with singular bath DOS with exponent $r_\mr{eff}=-r<0$.
In addition, we consider the particle-hole asymmetric critical fixed point,
for which we propose a universal field theory involving the crossing between
doublet and triplet levels.
\end{abstract}

\maketitle


\section{Introduction}

Quantum impurity models were introduced in the context
of dilute magnetic moments in metals and are based on the
notion that important aspects of the physics of extended systems
can be captured by spatially local processes. Indeed, the reduction of a
complicated many-body problem to the study of a {\it single} quantum degree of freedom
coupled to a simplified environment can be an important step in understanding
more complex physical situations.
Impurity models form the basis for the investigation of
lattice systems containing local moments -- these turn out to be extremely rich,
displaying various phenomena such as heavy-electron quantum coherence,
glassiness, and quantum phase transitions.
In addition, simpler yet interesting quantum effects that take place at
the level of a single magnetic moment can also be successfully described
using (oversimplified) single impurity models~\cite{hewson}.
The fermionic Kondo effect, i.e., the screening of an extra
magnetic moment by conduction electrons, is certainly the hallmark of this
single-impurity approach,
but other aspects like local non-Fermi liquid behavior and boundary quantum
phase transitions~\cite{vojta} have also been put forward.

In general, single-impurity models can serve as an excellent testing ground
for novel techniques and physical paradigms that
may be helpful in the more complicated case of extended or dense
impurity systems.
In this respect,
recent work based on the so-called pseudogap (or gapless) Kondo and Anderson models,
in which conduction electrons obey a semi-metallic density of states
that vanishes as $\rho(\w) \propto |\w|^r$ ($r>0$), has
shown interesting promises.
Indeed, this problem features a plethora of non-trivial properties, which can be
elegantly captured by {\it controlled} fermionic renormalization group (RG)
techniques~\cite{fradkin,fritz1}.
Powerful non-perturbative Numerical Renormalization Group (NRG)
simulations~\cite{fritz1,ingersent} were able to confirm the success
of the analytical approach.
In particular, recent work~\cite{fritz1,fritz2} has exposed that
all fixed points appearing in these pseudogap models can be
accessed perturbatively after identifying the value of the DOS
exponent $r$ that corresponds to their associated lower-critical or
upper-critical ``dimension''.

While the fixed points of the pseudogap spin-1/2 Kondo and single-orbital
Anderson models are now understood in this language,
here we are interested in certain extensions of these models
for which numerical data is also available:
the case of a spin-1 impurity has been investigated in the extensive work
of Gonzalez-Buxton and Ingersent~\cite{ingersent},
and has shown interesting yet unexplained properties.
We note that the physics of spin-1 Kondo models can be realized in
multilevel quantum dot
systems, and theoretical predictions for transport quantities
reflecting the singular Fermi liquid behavior of underscreened Kondo spins
in a metallic host \cite{singFL} have been put forward \cite{anna}.

In this paper, our objective is twofold:
first, we will generalize the theories for the critical fixed points,
which are already present in the spin-1/2 pseudogap Kondo model,
to the $S=1$ situation, following Refs.~\onlinecite{fradkin,fritz2}.
This includes two types of fixed points, one occuring at particule-hole (p-h)
symmetry and being controlled at small $r$, and the
other present away from p-h symmetry and captured near $r=1$  using a general
mapping onto an effective interacting resonant level model.
Second, we describe the physics of the {\it additional stable} fixed point
which emerges at particle-hole symmetry, due to the instability of the strong-coupling
symmetric fixed point for $S>1/2$.
Analyzing the leading relevant perturbation, we will be able to capture this
non-Fermi phase by a mapping onto an effective ferromagnetic
spin $S_\mr{eff}=1/2$ model with singular exponent
$r_\mr{eff}=-r<0$.
Technically, we will perform one-loop perturbative RG calculations together
with suitable $\epsilon$-expansions \cite{epsfoot}
to determine thermodynamic quantities;
the results compare favorably with the numerical data obtained in
Ref.~\onlinecite{ingersent}.
The present work also illustrates the broad applicability of the general
field-theoretic tools, developed in Refs.~\onlinecite{fradkin,fritz1,fritz2},
in a more complicated setting.

The bulk of the paper is organized as follows:
we start by presenting the spin-1 Kondo model with a given pseudogap
density of states (Sec.~\ref{sec:model}), and summarize previous results
from NRG calculations (Sec.~\ref{sec:num}).
In Sec.~\ref{sec:sym}, we focus on the case of particle-hole
symmetry.
It shows a critical point accessible from weak coupling
(which is a generalization to $S>1/2$ of the critical point analyzed by
Withoff and Fradkin \cite{fradkin}).
We also present an analysis and strong-coupling calculations for the
stable non-Fermi liquid fixed point found in the numerical simulations
of Ref.~\onlinecite{ingersent}.
Sec.~\ref{sec:asym} presents a novel field theory that
describes the quantum phase transition between the free-moment
and strong-coupling phases in the
presence of particle-hole asymmetry. In all cases, direct
comparison to NRG data is made.
Technical details on the diagrammatic approach can be found
in several appendices.
We will finally conclude the paper and discuss some open questions
related to the pseudogap Kondo model.


\section{The model}
\label{sec:model}

The pseudogap Kondo model (for a review, see Ref.~\onlinecite{vojta})
originates from the question of how a magnetic moment reacts
to the coupling to an electronic bath which has depleted weight
at the Fermi level, and is of interest to various condensed-matter
systems such as $d$-wave superconductors, graphite, and
semiconductors.  The basic physical idea is that the possibility
of observing a Kondo effect is weakened by the lack of low-energy
states, which results in various quantum phase transitions
controlled by the strength of the Kondo interaction, the
amount of particle-hole symmetry breaking, and obviously the shape of
the electronic density of states. We will model the latter by the
simplified form
\begin{equation}
\rho(\epsilon) = \sum_k \delta(\e-\ek) = N_0 |\e|^r \theta(D^2-\e^2) \,,
\label{dos}
\end{equation}
which has an algebraic dependence in energy characterized by the
number $r>0$ (the standard Kondo model corresponds to
$r=0$). Here $D$ is the half-bandwidth and $N_0=(r+1)/(2D^{r+1})$.

The Kondo model is characterized by an antiferromagnetic
coupling $\JB$ to this fermionic bath, and possibly a
potential scattering term $E_0$.
It reads
\begin{eqnarray}
\nonumber
{\cal H}_B & = & \sum_{k\s} \ek \cksd \cks + J_B \vec{S}_B \cdot \sum_{\s\s'}
\csd(\0) \frac{\vec{\tau}_{\s\s'}}{2} \csp(\0) \\
& & + E_0 \sum_{\s} \csd(\0) \cs(\0)
\label{kondo}
\end{eqnarray}
in standard notation,
and the subscript $_B$ stands for ``bare'' physical quantities and
operators, in view of the later RG calculations.
For $E_0 = 0$ the model is p-h symmetric, and $E_0$ tunes the amount of p-h
asymmetry.
In the following we will be concerned with the case of a spin-1
operator $\vec{S}_B$ and want to analyze the various fixed points of
this model.


\section{Summary of previous numerical results}
\label{sec:num}

\begin{figure}[!t]
\begin{center}
\includegraphics[width=8.5cm]{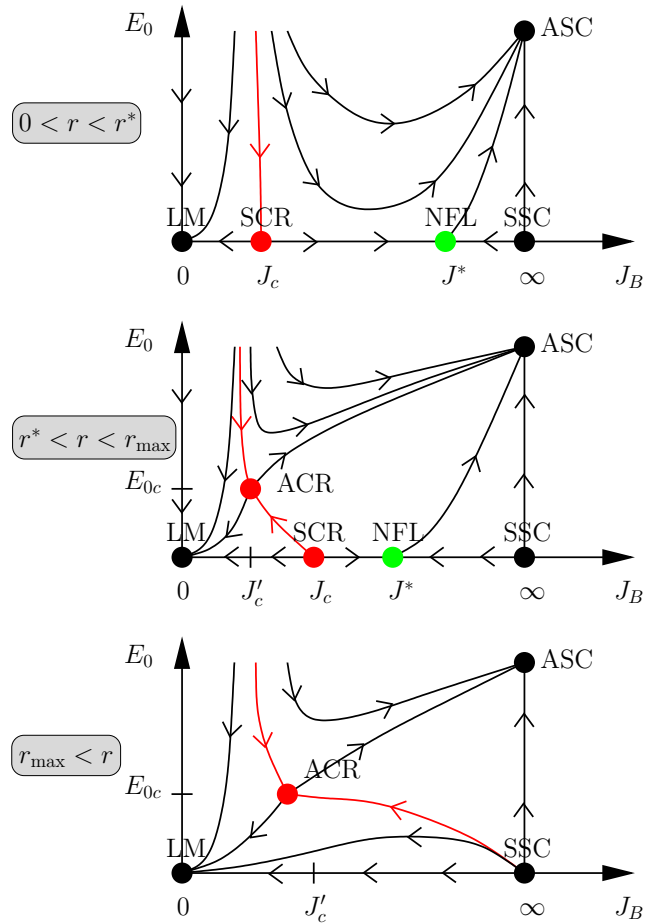}
\end{center}
\caption{
(color online)
Renormalization group flow diagrams for the $S=1$ pseudogap Kondo model,
deduced from NRG calculations~\cite{ingersent}.
The horizontal axis is the Kondo coupling, the vertical axis denotes the
amount of p-h symmetry breaking.
Depending on the value of the bath exponent $r$, three distinct situations are found.
The red lines represent phase boundaries, the red dots critical fixed points.
For a detailed explanation of the fixed points see text.}
\label{flow}
\end{figure}

The comprehensive numerical study of Gonzalez-Buxton
and Ingersent~\cite{ingersent} has exposed that the
model~(\ref{dos}-\ref{kondo}) displays several zero-temperature phases,
separated by boundary quantum phase transitions.
Their results are summarized in Fig.~\ref{flow} as a function
of Kondo coupling $\JB$, particle-hole asymmetry $E_0$,
and the value of the bath exponent $r$.

Let us focus on the p-h symmetric case, $E_0=0$, first. At $\JB=0$
we have a free spin $1$, dubbed local-moment (LM) fixed
point, which is stable w.r.t a small Kondo coupling $\JB$ for $r>0$.
Flow towards large coupling is possible only when
$\JB$ exceeds a critical value $J_c$ -- this phase transition
defines a symmetric critical (SCR) point.
Contrary to the $S=1/2$ case where a symmetric strong-coupling (SSC)
fixed point (corresponding to $\JB=\infty$) is always attained when $\JB>J_c$,
SSC is generically unstable in the $S=1$ model, which instead
shows an intermediate {\it stable} fixed point (NFL).
It is located at a finite value $J^*>J_c$ and corresponds to
a non-Fermi liquid (NFL) phase.
This flow structure does not persist for all values of $r$, since at
a value $r_\mr{max}\simeq 0.27$, the SCR and NFL fixed points
{\it collapse} and disappear altogether. For $r>r_\mr{max}$,
LM becomes then generically stable \cite{rmaxfoot}.

Now we turn to finite p-h asymmetry, i.e., potential scattering $E_0 \neq 0$.
Then SCR is stable against finite $E_0$ only for $0<r<r^*$ (where
$r^*\simeq0.245$), whereas NFL is always unstable. In the
case $r > r^*$, SCR becomes unstable as well, so that an asymmetric
cousin of the critical point, ACR, exists and controls the
quantum phase transition between LM and an asymmetric strong-coupling
(ASC) fixed point. ACR is associated to finite values
$J_c'$ and $E_{0c}$ of both the Kondo coupling and the
potential scattering term.

The numerical simulations of Ref.~\onlinecite{ingersent} have also
established the behavior of several observables,
such as the impurity contributions to the entropy and to the Curie
susceptibility, for the various fixed points. Whereas the
trivial cases (LM, SSC, ASC) are readily understood, all
intermediate-coupling fixed points (SCR, NFL, ACR) correspond to
{\it fractional} values of both the ground state degeneracy and
spin, and cannot be understood on the basis of an independent
electron picture.
The values extracted from the Numerical
Renormalization Group data are shown on Table~I, and in the
remainder of the paper we aim at an analytical calculation of
such singular properties, which characterize the complicated nature 
of the various ground states occuring in the model.

\begin{table}[ht]
\begin{center}
\begin{tabular}{|c|c|c|}
\hline
Fixed point $(J,E_0)$ & Entropy & Curie constant \\
\hline \hline
LM $(0,0)$ & $\ln 3$ & $2/3$\\
\hline
SCR $(J_c,0)$ & $\ln 3 + a r^3$ & $2/3 - 0.7 r$ \\
\hline
NFL $(J^*,0)$ & $\ln 2 +  2 r \ln 2$ & $1/4 + 0.4 r$ \\
\hline
SSC $(\infty,0)$ & $\ln 2 + 2r \ln 2$ & $1/4 + r/8$\\
\hline
ACR $(J_c',E_{0c})$ & $1.6 - 0.7 (1-r)$ & $0.5 - 0.04 (1-r)$ \\
\hline
ASC $(\infty,\infty)$ & $\ln 2$ & $1/4$ \\
\hline
\end{tabular}
\end{center}
\caption{
Numerical values of the impurity contribution to the entropy
and to the Curie constant, for the various fixed points of
the $S=1$ pseudogap Kondo model for $0\leq r \leq 1$,
as determined by NRG~\cite{ingersent}.
For LM, SSC, ASC the quoted values are exact.}
\end{table}


\section{Particle-hole symmetric model}
\label{sec:sym}

In this section we analyze the particle-hole symmetric spin-1
pseudogap Kondo model (\ref{kondo}), with $E_0=0$.
We start with the weak-coupling RG which allows to capture
the SCR fixed point for small $r$.
Then we consider an expansion around the symmetric strong-coupling
fixed point -- this will lead us to an effective ferromagnetic Kondo model
with singular density of states
which we use to describe the physics of the stable NFL fixed point.

\subsection{Weak-coupling analysis of the SCR fixed point}
\label{weak}

The study of the regime with small Kondo coupling and
particle-hole symmetry was done by Withoff and Fradkin~\cite{fradkin}
for $S=1/2$.
Perturbative RG is performed around $J_B=0$, i.e., the LM fixed point.
The tree level scaling dimension of the Kondo coupling is ${\rm dim}[J_B]=-r$,
and the RG flow in the vicinity of the LM fixed point can be described
by the one-loop beta function:
\begin{equation}
\beta(j) = r j - j^2 + \mcal{O}(j^3)
\label{betaj}
\end{equation}
where $j$ is the renormalized dimensionless Kondo coupling,
defined in Eq.~\ref{bareJ} below.
The beta function (\ref{betaj}) obviously yields an unstable fixed point
at $j_c = r$, which corresponds to SCR and controls the transition
between LM and NFL.
The correlation length exponent is
\begin{equation}
\frac{1}{\nu} = r + {\cal O}(r^2) \,,
\label{nuz_weak}
\end{equation}
and $r=0$ can be interpreted as lower-critical ``dimension'' for the
phase transition.

These results can be expected to persist for $S=1$, since the beta function for
the Kondo problem is known~\cite{fabrizio} to be independent of the
impurity spin value $S$.
However, as we are interested in computing physical observables at SCR,
which do differ from their $S=1/2$ counterparts,
we will proceed with a complete RG analysis for $S=1$ as well.
We only sketch here the intermediate results, while details on the derivation
can be found in Appendix~\ref{app1}.

\subsubsection{RG procedure}

The analysis is based on a representation of the spin 1
by a triplet of Abrikosov fermions, $f_{mB}$:
\begin{eqnarray}
S_B^+ & = & S_B^x+iS_B^y =  \sqrt{2} (f^\dagger_{1B} f^{\phantom{\dagger}}_{0B}
+f^\dagger_{0B} f^{\phantom{\dagger}}_{-1B}) \,, \nonumber \\
S_B^z & = & f^\dagger_{1B} f^{\phantom{\dagger}}_{1B} -
f^\dagger_{-1B} f^{\phantom{\dagger}}_{-1B}
\end{eqnarray}
where the Hilbert space constraint
\begin{equation}
Q = \sum_{m=1,0,-1} f^\dagger_{mB} f^{\phantom{\dagger}}_{mB} = 1
\end{equation}
is enforced \cite{kircan,lambda} using a chemical
potential $\lambda\to\infty$.

To proceed with the RG
scheme, we relate the bare coupling constant and field, $\JB$ and $f_{mB}$, to
renormalized quantities, $j$ and $f_m$
(which bear no index to shorten the notation within renormalized perturbation theory):
\begin{eqnarray}
\label{bareJ}
N_0 J_B & = & \mu^{-r} Z_f^{-1} Z_J j \,, \\
\label{baref}
f_{m B} & = & Z_f^{1/2} \fm
\end{eqnarray}
where $\mu$ is a renormalization energy scale.
These equations define renormalization factors $Z_J$ and $Z_f$
which will be determined using the standard field-theoretic RG
scheme \cite{bgz}, employing dimensional regularization and minimal
subtraction of poles.

A one-loop calculation (see Appendix~\ref{app1}) of the $\fm$
self-energy and the Kondo vertex yields
\begin{eqnarray}
\label{Zf}
Z_f & = & 1 + \mcal{O}(j^2) \,, \\
\label{ZJ}
Z_J & = & 1 + \frac{j}{r} + \mcal{O}(j^2) \,,
\end{eqnarray}
respectively.
The RG beta function is obtained by the condition
that the bare Kondo coupling is scale invariant, i.e.
$d J_B/d\mu = 0$, which gives from~(\ref{bareJ}):
\begin{equation}
\beta(j) \equiv \mu \frac{d j}{d\mu} = \frac{rj}
{1+j\frac{d}{d j} \ln(Z_J/Z_f)} \,.
\end{equation}
Inserting~(\ref{Zf})-(\ref{ZJ}) into this expression
proves that the beta function in the $S=1$ case
is still given by Eq.~(\ref{betaj}), as expected.
We also note that p-h asymmetry is irrelevant
at this fixed point, with $\beta(e_0) = r e_0$
(where $e_0 \propto N_0 E_0$ is the dimensionless potential
scattering term).

\subsubsection{Impurity entropy}

The impurity contribution to the entropy is obtained as
the difference of the total entropy and the entropy of the
electron bath alone.
Its $T=0$ value represents a measure of the impurity ground
state degeneracy.
For the SCR fixed point it turns out that
corrections to the LM value are extremely small \cite{kircan}:
\begin{equation}
\Simp^{\mr{SCR}} = \ln 3 + \mcal{O}(r^3)~.
\end{equation}
This makes the comparison to numerical data difficult, and
we refrain from extracting the prefactor of the $r^3$ term.

\subsubsection{Impurity susceptibility}

The impurity susceptibility $\Cimp$ is similarly calculated
by removing the free (i.e. $J_B=0$) contribution of the bulk
fermions to the total spin susceptibility \cite{suscfoot},
\begin{equation}
\label{chitot}
\chi_\mr{tot}(T)  =  \inte \big< S^z_\mr{tot}(\tau)
S^z_\mr{tot}(0)\big> \,,
\end{equation}
with
\begin{equation}
\label{Stot}
S^z_\mr{tot}  =  \sum_m m f^\dagger_{m B} f^{\phantom\dagger}_{m B} +
\sum_{\x\s} \frac{(-1)^\s}{2} \csd(\x) \cs(\x) \,.
\end{equation}
Apart from numerical factors, the calculation is analogous to the
$S=1/2$ case~\cite{kircan}, and we find at lowest order in $j$:
\begin{equation}
\label{Cimpweak}
\Cimp = \frac{2}{3T} - j \frac{2}{3T}
\left(\frac{\mu}{T}\right)^{-r}
\int_{-\infty}^{+\infty} \mr{d}x
\; \frac{e^{-x}}{(e^{-x}+1)^2}
\end{equation}
Because we want to obtain the SCR fixed point contribution, we can
replace $j=j_c=r$ and take $(\mu/T)^{-r} \simeq 1$ as
$r\rightarrow0$, which gives finally the effective Curie constant:
\begin{equation}
T \Cimp^\mr{SCR} = \frac{2}{3} - \frac{2r}{3} \,,
\end{equation}
in agreement with the numerical value found in Table~I.

\subsection{Non-Fermi liquid phase near strong coupling}

\subsubsection{Around the limit $J=\infty$}

To understand the emergence of a stable non-trivial fixed
point, we consider in the first place the vicinity of the
SSC fixed point.
Indeed, the renormalization flow sketched from the
numerical solution demonstrates that the strong-coupling
fixed point is unstable, which is quite curious at first
thought.
Indeed, at large $J$, the spin 1 is partially
screened by the fermions at site $\0$, leaving a remnant
effective spin $S_\mr{eff}=1/2$.
Applying the Nozi\`eres-Blandin argument~\cite{nozieres}
we see that an effective ferromagnetic coupling $J_\mr{eff} = - t_{01}^2/J$
is then generated between $\vec{S}_\mr{eff}$ and the
electrons $\csd(\1)$ sitting on the site near $\csd(\0)$
(above, $t_{01}$ is the typical hopping between $\csd(\0)$ and
$\csd(\1)$, and we assumed a chain representation of the
electronic bath).
Naively we would then expect SSC to be stable.
This is {\em not} the case, because at particle-hole
symmetry the Green's functions of the above electrons are
related through:
\begin{equation}
\big<\csd(\0,i\w)\cs(\0,i\w)\big>^{-1} = i\w -
t_{01}^2 \big<\csd(\1,i\w)\cs(\1,i\w)\big>
\end{equation}
which shows that the fermions $\csd(\1)$ have an effective density
of states $\rho_\mr{eff}(\e) \propto |\e|^{r_\mr{eff}}$, with
$r_\mr{eff} = -r <0$.

Thus, the effective Hamiltonian near SSC is a ferromagnetic
spin-1/2 Kondo model with a {\it negative} DOS exponent.
Interestingly, this model was studied before in Ref.~\onlinecite{bulla},
and was found to display a stable intermediate-coupling fixed point.
To see this, we note that
the beta function for $J_\mr{eff}$ reads
\begin{equation}
\beta(j_\mr{eff}) = r_\mr{eff} j_\mr{eff}  - j_\mr{eff} ^2 =
-r j_\mr{eff}  - j_\mr{eff} ^2 \,.
\end{equation}
This yields a stable fixed point at $j_\mr{eff}^* = -r$,
which is allowed since the effective coupling is
ferromagnetic! In our original language, this is the NFL
fixed point, which controls now a whole phase with non-trivial
properties, and is a unique feature of the $S>1/2$
pseudogap Kondo model. In particular, the above arguments
show that this fixed point is located at $j^* \sim 1/r$.
The NFL fixed point is unstable with respect to
particle-hole asymmetry, with $\beta(e_0) = - r e_0$,
a situation which will be discussed in Sec.~\ref{sec:asym}.
Finally, we note that the values for $r^*$ and $r_\mr{max}$
obtained by the Numerical Renormalization Group for the
original problem~\cite{ingersent} and for the effective
model~\cite{bulla} coincide, in agreement with our
identification.
We would like now to compute the physical
quantities at the intermediate NFL fixed point.

\subsubsection{Impurity entropy}

We know that the correction to the entropy at a
particle-hole symmetric Withoff-Fradkin type of fixed point
is as small as $r^3$. However, we are now expanding around a
strong-coupling fixed point and not with respect to a free
spin $S=1$. Since the impurity entropy is defined relative to the
decoupled ($J=0$) limit, we have to take into account the
contribution at SSC ($\JB=\infty$), which contains a $\ln 2$
term due to the underscreened spin $S_\mr{eff}=1/2$, and
depends on $r$ as well~\cite{ingersent,fritz1}, due to the
singular nature of the resonant level limit (although the
SSC fixed point is trivial). We obtain finally
\begin{equation}
\Simp^\mr{NFL} = \ln 2 + 2 r \ln 2 + \mcal{O}(r^3) \,,
\end{equation}
as anticipated numerically (Table~I).

\subsubsection{Impurity susceptibility}

We have to follow the same argument as above, and sum the
contribution of the strong-coupling fixed point,
$T\Cimp^\mr{SSC} = 1/4+r/8$ (see~\cite{fritz1}) as well as the
non-trivial correction due to the stable $S_\mr{eff}=1/2$
ferromagnetic fixed point with exponent $r_\mr{eff}=-r$,
which is~\cite{kircan} $\Delta(T\Cimp) = -r_\mr{eff}/4 = r/4$.
The complete result thus agrees with the value in Table~I:
\begin{equation}
T\Cimp^\mr{NFL} = \frac{1}{4}+\frac{3r}{8} \,.
\end{equation}


\section{Particle-hole asymmetric model}
\label{sec:asym}

The nature of the particle-hole asymmetric fixed point
of the pseudogap Kondo model, present at $E_0\neq0$ and $r > r^*$,
was understood in the recent work~\cite{fritz1,fritz2}.
The crucial point noticed there is that this quantum
phase transition separates the local-moment fixed point from
the asymmetric strong-coupling one, as can be checked in
Fig.~\ref{flow}, so that the corresponding field theory
can be constructed by ``mixing'' the relevant degrees of freedom
associated to both fixed points.
In the $S=1/2$ case, one needed to consider a level crossing
of a doublet (associated to LM) and a singlet (from to ASC),
with transitions allowed through the coupling to the conduction
electrons.
Not surprisingly, the effective theory took the form of an infinite-$U$,
i.e., maximally p-h asymmetric, Anderson model.
This was perfectly consistent with numerical data which
indicate the phase transitions in the pseudogap Kondo and Anderson
models are in the same universality class.
The theory could be analyzed by expanding around the level-crossing
(valence fluctuation) fixed point, and it was shown that
$r=1$ plays the role of an upper-critical ``dimension''.

In the following, we want to generalize the idea of Refs.~\onlinecite{fritz1,fritz2}
to the spin-1 case.

\subsection{Derivation of the critical field theory}

In the present problem, the spin $S=1$ will be described
by a triplet of fermions $\fm$ near LM. On the other hand, ASC
shows underscreening of the spin, and gives a remnant spin-1/2
moment, that we can describe by a doublet of bosons $\bs$.
Having now the important variables to describe the
transition at ACR, we need to write down the effective theory.
Several requirements have to be imposed: the Hamiltonian
should obey spin rotation invariance and should reduce to a
spin 1 (resp. 1/2) Kondo Hamiltonian near the LM (resp. ASC) limit.
One naturally arrives at the following generalization
of the infinite-$U$ Anderson model of Ref.~\onlinecite{fritz1}:
\begin{eqnarray}
\label{heff}
{\cal H} & = & \sum_{k\s} \ek \cksd \cks + \epsilon_f \sum_m \fmd
\fm \\
\nonumber
& +& V_B \big[f^\dagger_{0B} b^{\phantom{\dagger}}_{\downarrow B} \,
c^{\phantom{\dagger}}_\uparrow(\0) +
f^\dagger_{0B} b^{\phantom{\dagger}}_{\uparrow B} \,
c^{\phantom{\dagger}}_\downarrow(\0) \big] + \mr{h.c.} \\
\nonumber
& + & \sqrt{2} V_B \big[
f^\dagger_{1 B} b^{\phantom{\dagger}}_{\uparrow B} \, c^{\phantom{\dagger}}_\uparrow(\0) +
f^\dagger_{-1 B} b^{\phantom{\dagger}}_{\downarrow B} \, c^{\phantom{\dagger}}_\downarrow(\0)
\big] + \mr{h.c.} \,.
\end{eqnarray}
Here, $\epsilon_f$ is the parameter used to tune the system through
the phase transition; at tree level the transition occurs at
$\epsilon_f=0$, and corresponds to the crossing of doublet and triplet states,
which we call the valence fluctuation (VFl) fixed point.
The restriction to the physical Hilbert space is implemented with
the constraint
\begin{equation}
Q= \sum_m f^\dagger_{m B} f^{\phantom\dagger}_{m B} +
\sum_\s b^\dagger_{\sigma B} b^{\phantom\dagger}_{\sigma B} = 1 \,.
\end{equation}
One can check easily via a Schrieffer-Wolff transformation
that ${\cal H}$ (\ref{heff}) evolves onto a spin 1 (resp. 1/2) Kondo model when
$\epsilon_f$ is large and negative (resp. positive), as it
should be.

The effective Hamiltonian~(\ref{heff}) seems to have nothing
to do with the original model~(\ref{kondo}), and in particular it is not
as simply understandable as in the $S=1/2$ case considered
in~\cite{fritz2}.
However, we will show that it correctly describes the critical
properties at ACR.
A number of facts are encouraging:
Power counting at the $\epsilon_f=V_B=0$ fixed point shows that
the coupling constant $V_B$ has a tree-level scaling dimension
${\rm dim}[V_B]=\bar{r}=(1-r)/2$, i.e., it is relevant for $r<1$,
which indicates the possibility of a
non-trivial critical point in this regime.
In contrast, $V_B$ is irrelevant and will flow to zero for $r>1$,
which establishes the role of $r=1$ as upper-critical dimension
where $V_B$ is marginal.
Similar to the situation in the spin-1/2 model,
the transition will become a level crossing with perturbative corrections
for $r>1$.
For $V_B \to 0$ we have strong ``valence'' fluctuations between the
doublet and triplet levels,
leading to an entropy of $\ln 5$, in agreement with the numerical
result shown in Table~I.

Motivated by this we proceed with an RG expansion around the decoupled
$V_B=0$ fixed point.
As demonstrated below, we find a flow diagram
identical in structure to the one of the spin-1/2 model,
Fig.~1 of Ref.~\onlinecite{fritz2}, with the difference that
the VFl fixed point now corresponds to a crossing of doublet and triplet
levels.
This flow diagram is very similar to the one of a standard $\phi^4$ theory.
In particular, the non-trivial critical point occuring for $r<1$
(which can be understood as the analogue of the Wilson-Fisher fixed point)
will be identified with the asymmetric critical point
(ACR) of the original Kondo Hamiltonian.


\subsection{RG calculation for $r$ close to 1}
\label{infu}

\subsubsection{RG procedure}

The renormalization method developed in Ref.~\onlinecite{fritz2} differs slightly from
the technique used in Sec.~\ref{weak} and Appendix~\ref{app1}.
We still need to introduce renormalization factors for the hybridization
strength and fields:
\begin{eqnarray}
N_0 V_B & = & \mu^{+\bar{r}} Z_f^{-1/2} Z_b^{-1/2} Z_V v \,, \\
f_{m B} & = & Z_f^{1/2} \fm \,, \\
b_{\s B} & = & Z_b^{1/2} \bs \,,
\end{eqnarray}
as well as mass renormalization terms,
\begin{equation}
{\cal H}_\lambda = \delta \lambda_f \sum_m \fmd \fm
+\delta \lambda_b \sum_\s \bsd \bs \,,
\end{equation}
already written using renormalized quantities.
We have introduced again a running renormalization scale
$\mu$ and used $\bar{r} = (1-r)/2$.

The RG will be performed at criticality, i.e., the $\delta\lambda$ are
chosen such to cancel the real parts of the bare self-energies.
Including the contribution of the counter-terms associated
to the above renormalization constants, we find (see
Appendix~\ref{app2}) the $f$-electron self-energy contribution
at one-loop:
\begin{equation}
\label{Sf}
\Sigma_f(i\w) = - 2 v^2 \left(\frac{\mu}{D}\right)^{\bar{r}}
\!\! \left[D\!+\!\frac{i\w}{2\bar{r}}
\left(\frac{D}{|\w|}\right)^{\bar{r}}\right]
+ \delta \lambda_f -i\w(Z_f-1) \,.
\end{equation}

By definition, the field renormalization is determined by
the condition $(\partial/\partial\w) \Sigma_f(0) = 0$, and
criticality is ensured by the vanishing of the mass,
$\Sigma_f(0)=0$, which gives:
\begin{eqnarray}
Z_f & = & 1 - \frac{v^2}{\bar{r}} + \mcal{O}(v^4) \,, \\
\label{dlf}
\delta \lambda_f & = & 2 v^2 \left(\frac{\mu}{D}\right)^{\bar{r}} D
+ \mcal{O}(v^2) \,.
\end{eqnarray}
The bosonic self-energy is similarly calculated (only
numerical prefactors differ) and leads to:
\begin{eqnarray}
Z_b & = & 1 - \frac{3v^2}{2\bar{r}} + \mcal{O}(v^4) \,, \\
\label{dlb}
\delta \lambda_b & = & 3 v^2 \left(\frac{\mu}{D}\right)^{\bar{r}} D
+ \mcal{O}(v^2) \,.
\end{eqnarray}
At this order, the hybridization is left unrenormalized, $Z_V=1$.
Finally, imposing the scale invariance condition
$\mr{d}V_B/\mr{d}\mu=0$, we find the beta function:
\begin{eqnarray}
\beta(v) & \equiv & \mu \frac{\mr{d}v}{\mr{d}\mu} =
\frac{-\bar{r}v}
{1+v\frac{d}{d v} \ln(Z_V Z_f^{-1/2} Z_b^{-1/2})} \nonumber\\
& = & -\bar{r}v + \frac{5}{2}v^3 + \mcal{O}(v^5).
\end{eqnarray}
This gives the fixed point value $(v^*)^2 = (2/5)\bar{r}$.
The correlation length exponent of the LM--ASC transition
can be obtained similar to
Ref.~\onlinecite{fritz2}, with the result
\begin{equation}
\frac{1}{\nu} = r + {\cal O}(\bar{r}^2) ~~~~(r<1) \,,
\label{nuz_infu}
\end{equation}
whereas a level crossing (formally $\nu = 1$) occurs
for $r>1$.
The RG flow is thus identical to Fig.~1 of Ref.~\onlinecite{fritz2}.

\subsubsection{Impurity entropy}

The correction at order $v^2$ to the free energy is
evaluated in Appendix~\ref{app2} and reads:
\begin{equation}
\label{DeltaF}
\Delta F = - \frac{12}{5} v^2 \mu^{2\bar{r}}
\int_{0}^D \mr{d}\e \; \frac{\e^r}{\e} \; \tanh\frac{\e}{2T}
\end{equation}
Using $\Delta \Simp = - \partial (\Delta F) /\partial T$, and
taking the limit $\bar{r}\rightarrow0$, we obtain finally:
\begin{equation}
\Simp^\mr{ACR} = \ln 5 -\frac{24\ln 2 }{25} (1-r)
\end{equation}
which is a agreement with the result in Table~I.

\subsubsection{Impurity susceptibility}

The computation of the Curie constant is done along the same
line, but is more involved due to the proliferation of
Feynman graphs, since the total spin operator
\begin{eqnarray}
S^z_\mr{tot}
&=& \sum_m m f^\dagger_{m B} f^{\phantom\dagger}_{m B} \nonumber \\
&+& \sum_\s
\frac{(-1)^\s}{2} \left[ b^\dagger_{\sigma B} b^{\phantom\dagger}_{\sigma B} +
\sum_\x \csd(\x) \cs(\x) \right]
\end{eqnarray}
mixes all three possible fields in the calculation.
We simply quote the final result, derived in
Appendix~\ref{app2}:
\begin{eqnarray}
\nonumber
\Cimp & = & \frac{1}{2T} - \frac{v^2\mu^{2\bar{r}}}{10 T}
\int_{0}^D \mr{d}\e \; \frac{\e^r}{T^2} \; \left[
\frac{\sinh \e/T}{1+\cosh \e/T}-1
\right. \\
& & \left.  + \frac{6\sinh \e/T}{(1+\cosh \e/T)^2} \right] .
\end{eqnarray}
We finally insert the fixed-point value $v^*$ and take the
limit $\bar{r}\rightarrow0$, which gives:
\begin{equation}
T\Cimp^\mr{SCR} = \frac{1}{2} - \frac{3-2\ln2}{50} (1-r) \,.
\label{cscr}
\end{equation}
Note that the leading Curie constant above is associated to the trivial 
susceptibility of a magnetic system that consists of a triplet
and a singlet. Because of numerical prefactors, e.g. in the partition
function, it is not simply given by the expression valid for a triplet
state only, namely S(S+1)/3=2/3, but by the different value 1/2.
Finally, we check that equation~(\ref{cscr}) coincides with the results in Table~I and
vindicates the usefulness of the effective
theory~(\ref{heff}) in describing the ACR fixed point.


\section{Conclusion}

In this paper, we have studied the spin-1 pseudogap Kondo
model using perturbative RG together with renormalized perturbation theory.
We have shown that all non-trivial fixed points are accessible in the vicinity
of their associated upper-critical or lower-critical dimension.
In particular, we have proposed a novel field theory
governing the critical behavior in the presence of
particle-hole asymmetry, and we also have provided a physical picture
for the stable particle-hole symmetric non-Fermi liquid phase
(which is not present in the $S=1/2$ case) using a strong-coupling
mapping to an effective ferromagnetic spin-1/2 model with singular
density of states.
The generic instability of the symmetric strong-coupling fixed point SSC
and the emergence of the non-Fermi liquid phase presents a nice example
of a situation where the purely local strong-coupling picture of
Kondo singlet formation is clearly inappropriate.
We also note that, from the arguments given above, our qualitative results are
expected to be generic to the pseudogap Kondo model for all spin values $S$
greater or equal to 1.

\begin{figure}[!t]
\begin{center}
\includegraphics[width=8.9cm]{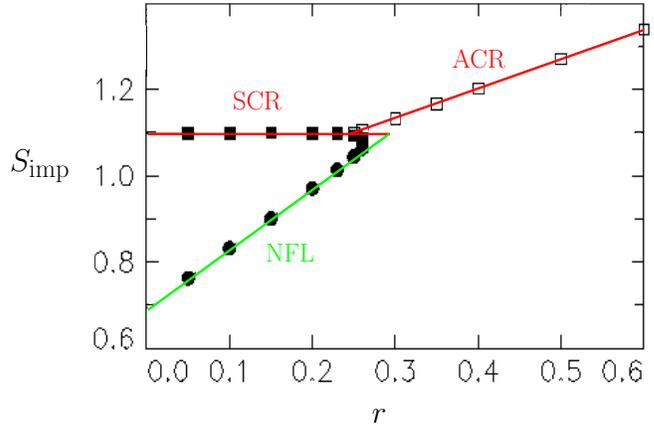}
\end{center}
\caption{
(color online) Comparison of the impurity entropy $S_{\rm imp}$ between the analytical
estimates at one-loop order (continuous lines) and the NRG data of
Ref.~\onlinecite{ingersent} (dots and squares) for the three different
intermediate-coupling fixed points occuring in the $S=1$ pseudogap Kondo model.
}
\label{comp_Simp}
\end{figure}

\begin{figure}[!t]
\begin{center}
\includegraphics[width=8.9cm]{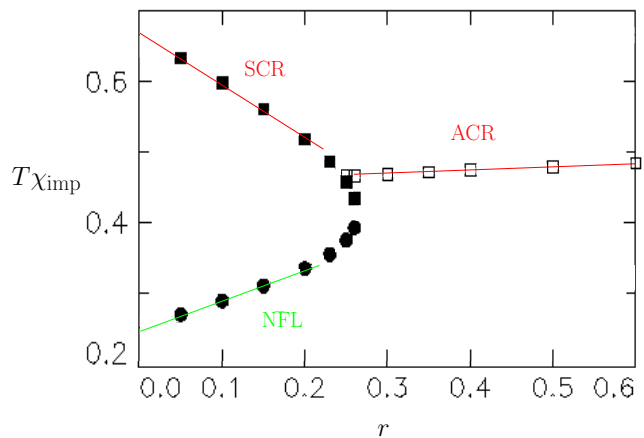}
\end{center}
\caption{
(color online) As Fig.~\protect\ref{comp_Simp}, but for the effective Curie constant
$T\chi_{\rm imp}$.
}
\label{comp_Cimp}
\end{figure}

The collected analytical results for thermodynamic quantities established in this paper
can finally be compared
to the numerical values obtained from the simulations~\cite{ingersent},
and are displayed in Fig.~\ref{comp_Simp} and \ref{comp_Cimp}.
The agreement is excellent, even at one-loop order,
demonstrating the usefulness of $\epsilon$-expansions for impurity models with
power-law bath spectra.


Using the present RG methods other observables can be calculated as well \cite{fritz2},
like the critical exponents for the local susceptibility or the
conduction electron $T$ matrix.
We only mention here the result for the $T$ matrix: similar to the spin-1/2
model the critical exponent can be determined exactly to all orders in
perturbation theory, with the result
\begin{equation}
{\rm Im} \, T(\omega) \propto \frac{1}{\omega^r} ~~~(0<r<1)
\end{equation}
which holds at all intermediate coupling fixed points (SCR, NFL, ACR).
At the critical dimensions, $r=0$ and $r=1$, logarithmic corrections
to this result occur.

The application of our results to experiments requires host systems
with pseudogap DOS.
Candidates are $d$-wave superconductors or materials with a special
semiconducting band structure (as planar graphite) --
these obey a linearly vanishing density of states.
Systems with exponent $0<r<1$ are difficult to realize;
proposals have been made e.g. using disordered mesoscopic conductors
\cite{hopk}.

On the theoretical side, the single-impurity pseudogap Kondo problems
are now reasonably well understood.
One open issue is a possible analytical description of a collapse of
intermediate-coupling fixed points, which occurs both as $r\to {r^*}^+$
(where two critical fixed points merge) and at $r\to {r_\mr{max}}^-$
(where the stable and the critical p-h symmetric fixed points
meet and disappear).
One idea comes to mind
when comparing the entropies for a spin-$S$ model:
the critical fixed point has $\Simp(J_c) = \ln(2S+1)+\mcal{O}(r^3)$,
whereas for the stable NFL fixed point has
$\Simp(J^*) = \ln(2S) + 2r \ln 2+\mcal{O}(r^3)$.
The two values cross at $r_\mr{max} \sim 1/(S \ln 8)$, a value that
becomes small if $S\gg1$.
One may hope that a controlled weak-coupling theory
describing the physics of $r_\mr{max}$ or $r^\ast$ exists
for large $S$, although this remains to be demonstrated.
Finally, we point out that generalizations of the pseudogap
model to several impurities can show an interesting
interplay of critical points with different nature.

\acknowledgments

We thank L. Fritz and M. Kir\'can for useful discussions.
This research was supported by the DFG Center for Functional Nanostructures
and the Virtual Quantum Phase Transitions Institute in Karlsruhe.


\appendix
\section{Weak-coupling RG}
\label{app1}

Here we provide calculational details for the weak-coupling RG
of Sec.~\ref{weak}.

\subsection{Vertex renormalization at one loop}

\begin{figure}[!t]
\begin{center}
\includegraphics[width=7cm]{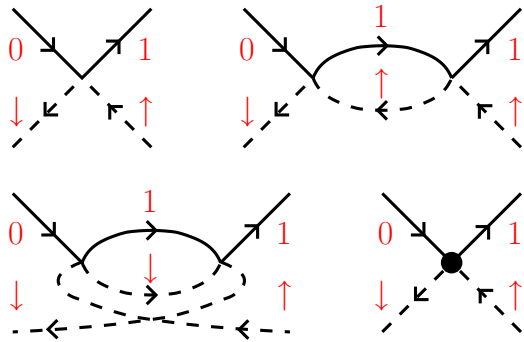}
\end{center}
\caption{(color online). Renormalization
of the vertex $\Gamma_{10\downarrow\uparrow}^{(1)}$
at one-loop level.
Continuous and dashed lines denote the free Green's functions of the
$f_m$ and $c_\sigma$ fields respectively, and the dot represents
the vertex counter-term $(Z_J-1)\Gamma_{10\downarrow\uparrow}^{(0)}$.
}
\label{gamma}
\end{figure}

We first rewrite the bare Kondo Hamiltonian~(\ref{kondo}) in
terms of renormalized quantities, allowing to separate
the bare action $\mcal{S}_B$ into a renormalized Kondo part
$\mcal{S}$, a constraint term $\mcal{S}_\lambda$, and a series
of counter-terms $\delta \mcal{S}$:
\begin{eqnarray}
\mcal{S}_B & = & \mcal{S} + \mcal{S}_\lambda + \delta
\mcal{S} \,, \\
\nonumber
\mcal{S} & = & \inte \sum_k \cksb (\dt + \ek) \cks
+ \sum_m \fmb \dt \fm  \\
& & + \inte \!\! \sum_{mm'\s\s'} \Gamma_{mm'\s\s'}^{(0)} \fmb \fm
\csb(\0)\cs(\0) \,, \\
\mcal{S}_\lambda & = & \inte \; \lambda \bigg(\sum_m \fmb \fm - 1\bigg) \,, \\
\delta \mcal{S} & = & \inte \; \delta\lambda \sum_{m} \fmb \fm \\
\nonumber
& & + \inte \; (Z_J-1) \!\! \sum_{mm'\s\s'}
\Gamma_{mm'\s\s'}^{(0)} \fmb \fm \csb(\0)\cs(\0) \,.
\end{eqnarray}
With a renormalized dimensionful coupling $J = j/N_0$
the tree level vertices satisfy
$\Gamma_{10\downarrow\uparrow}^{(0)} =
\Gamma_{0-1\downarrow\uparrow}^{(0)}  =
\Gamma_{01\uparrow\downarrow}^{(0)}  =
\Gamma_{-10\uparrow\downarrow}^{(0)}  = \mu^{-r} J/\sqrt{2}$ and
$\Gamma_{mm\s\s}^{(0)}  = \mu^{-r} J m (-1)^{\s}/2$, with all other
terms being zero.
Indeed, inserting the definitions~(\ref{bareJ})-(\ref{baref}) into
the above expression for the action, one recovers the bare action
of the problem. The free propagators are given by:
\begin{eqnarray}
G_{c0}(i\w,k) & = & \frac{1}{i\w-\ek} \\
G_{f0}(i\w) & = & \frac{1}{i\w-\lambda}
\end{eqnarray}
where the limit $\lambda\rightarrow\infty$ should be taken
only at the end of the calculation \cite{lambda}.
The diagrammatic expansion at one loop of the vertex
$\Gamma_{10\downarrow\uparrow}$ is given in
Fig.~\ref{gamma}.

Evaluating the loop at zero temperature, one finds:
\begin{eqnarray}
\nonumber
\Gamma_{10\downarrow\uparrow}^{(1)}(i\nu) & = &
\mu^{-r} Z_J \frac{J}{\sqrt{2}}
-\mu^{-2r}\frac{J^2}{\sqrt{2}} \int_{-D}^0 \mr{d}\e \; N_0
\frac{|\e|^r}{\e+i\nu} \\
& = & \frac{J}{\sqrt{2}} \mu^{-r} \left[ Z_J -
\frac{j}{r}(1+r\ln\big(|\nu|/\mu)\big) \right]
\end{eqnarray}
where we have used $j = N_0 J$, and the limit $r\rightarrow0$ has
been taken in the last integral.
The renormalization condition for the vertex is
$\Gamma_{10\downarrow\uparrow}^{(1)}(0) =
\Gamma_{10\downarrow\uparrow}^{(0)} = \mu^{-r}J/\sqrt{2}$, so
that the result~(\ref{ZJ}) follows.

\subsection{Impurity susceptibility}

\begin{figure}[!b]
\begin{center}
\includegraphics[width=8.4cm]{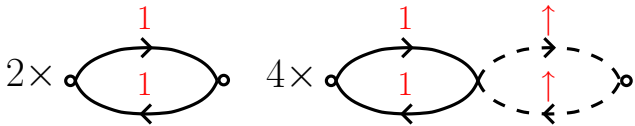}
\end{center}
\caption{(color online). Diagrammatic expansion of
$\big<Q\big> \Cimp$ at lowest order for the Kondo model at
SCR, including the complete prefactors.}
\label{chiweak}
\end{figure}

The diagrammatic expansion of the impurity susceptibility
follows from its definition~(\ref{chitot}-\ref{Stot}), and
accounting for all numerical prefactors (which arise either
from the $(-1)^\s/2$ term in~(\ref{Stot}) or from combinatorial
reasons), we have to calculate the contribution shown in
Fig.~\ref{chiweak}. Note that $\big<Q\big> \Cimp$ and not
$\Cimp$ has a simple diagrammatic expansion, since the
constraint $Q= \sum_m f^\dagger_{m B} f^{\phantom\dagger}_{m B} =1$  means that we are {\em not}
expanding around a free-particle limit. At lowest
order, we have $\big<Q\big> = 3 \exp(-\beta\lambda)$.
Computing the Matsubara sums at finite temperature, one gets
the result~(\ref{Cimpweak}) in the limit where the
half-bandwidth $D$ goes to infinity.

\section{RG around the valence-fluctuation limit}
\label{app2}

This appendix contains details for the RG calculation
of Sec.~\ref{infu}.

\subsection{Pseudoparticle self-energies at one loop}

\begin{figure}[!t]
\begin{center}
\includegraphics[width=8.6cm]{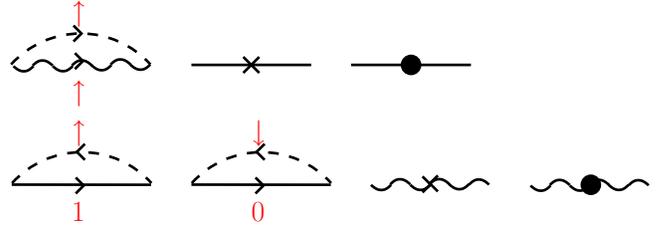}
\end{center}
\caption{(color online). Self-energy at one loop for the
$f_1$ fermion (upper part) and $b_{\uparrow}$ boson
(lower part) respectively. Wiggly lines represent the
bosonic free propagators. Crosses and dots indicate
respectively mass and field renormalization counter-terms.}
\label{self}
\end{figure}

Again, we will set up notations and express
the bare action
$\mcal{S}_B = \mcal{S} + \mcal{S}_\lambda + \delta \mcal{S}$ in terms
of renormalized quantities:
\begin{eqnarray}
\mcal{S} & = & \inte \sum_{k\s} \cksb (\dt+\ek) \cks \\
\nonumber
& & + \inte \sum_m \fmb (\dt + \epsilon_f) \fm + \sum_\s \bsb \dt \bs\\
\nonumber
& & + \inte  V \mu^{\bar{r}} \big[{\bar f}_0 b^{\phantom{\dagger}}_\downarrow
c^{\phantom{\dagger}}_\uparrow(\0) +
{\bar f}_0 b^{\phantom{\dagger}}_\uparrow
c^{\phantom{\dagger}}_\downarrow(\0) \big] + \mr{c.c.} \,, \\
\nonumber
& & + \inte \sqrt{2} V \mu^{\bar{r}} \big[
{\bar f}_1 b^{\phantom{\dagger}}_\uparrow c^{\phantom{\dagger}}_\uparrow(\0) +
{\bar f}_{-1} b^{\phantom{\dagger}}_\downarrow c^{\phantom{\dagger}}_\downarrow(\0)
\big] + \mr{c.c.} \,, \\
\mcal{S}_\lambda & = & \inte \; \lambda \bigg(\sum_m \fmb \fm
+\sum_\s \bsb \bs  - 1\bigg) \,, \\
\delta \mcal{S} & = & \inte \; \delta\lambda_f \sum_{m} \fmb \fm
+ \delta\lambda_b \sum_{\s} \bsb \bs + \ldots
\end{eqnarray}
As above, we have introduced a renormalized dimensionful hybridization $V = v/N_0$.
We did not write the vertex renormalization factor, since it
will not appear at this order of the calculation. We also
have to introduce the bosonic Green's function~\footnote{Because
of the minus sign involved in this definition, an extra $(-1)$
factor should be associated to each bosonic lines in the diagrammatics.}:
\begin{equation}
G_{b0}(i\nu) = - \big<\bsd(i\nu) \bs(i\nu)\big> = \frac{1}{i\nu-\lambda} \,.
\end{equation}
Calculating the self-energy for the $f_1$ fermion
shown in Fig.~\ref{self}, we find at zero temperature:
\begin{equation}
\Sigma_f(i\w) = - 2 N_0 V^2 \mu^{2\bar{r}} \int_0^D \mr{d}\e
\frac{\e^r}{\e-i\w} + \delta\lambda_f -i\w(Z_f-1)
\end{equation}
which reduces to~(\ref{Sf}) when the limit
$\bar{r}\rightarrow0$ is taken. The calculation for the
$b_\uparrow$ proceeds along the same line, although
the prefactors differ slightly (see Fig.~\ref{self}).

\subsection{Impurity entropy}

\begin{figure}[!b]
\begin{center}
\includegraphics[width=6.5cm]{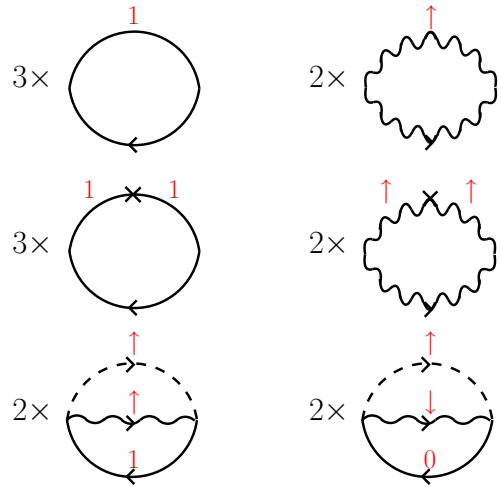}
\end{center}
\caption{(color online). Expansion of the partition function
$\mcal{Z}$ up to order $V^2$ and $\delta\lambda_{f,b}$, including
the combinatorial factors.}
\label{part}
\end{figure}

To evaluate the impurity entropy, we need to expand the
physical partition function $\mcal{Z}$, which is represented
diagrammatically in Fig.~\ref{part}. The free energy can
then be obtained through $F = -T\ln \mcal{Z}$, and reads:
\begin{eqnarray}
\nonumber
F & = & -T \ln 5  + \frac{12}{5} v^2 \mu^{2\bar{r}}
\int_{0}^D \mr{d}\e \; \frac{\e^r}{\e} \\
& & - \frac{12}{5} v^2 \mu^{2\bar{r}}
\int_{0}^D \mr{d}\e \; \frac{\e^r}{\e} \; \tanh\frac{\e}{2T}
\label{F_ACR}
\end{eqnarray}
which is the result quoted in Eq.~(\ref{DeltaF}) from
which the impurity entropy at the ACR fixed point follows.
In deriving this result we have used the renormalized values
for the counter-terms~(\ref{dlf})-(\ref{dlb}), which provide
the correction in the first line of the above expresion.
One can check that
they actually drop out when the entropy is calculated, but those
corrections are nevertheless crucial for regularizing the
forthcoming impurity susceptibility.

\subsection{Impurity susceptibility}

\begin{figure}[!t]
\begin{center}
\includegraphics[width=8.7cm]{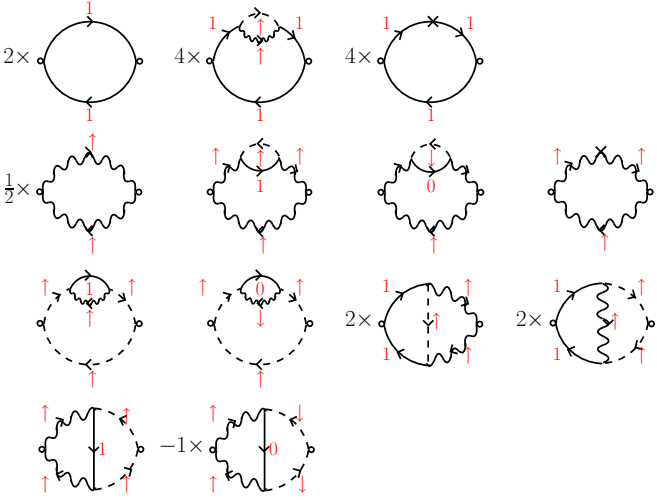}
\end{center}
\caption{(color online). Diagrammatic expansion of $\big<Q\big> \Cimp$
at lowest order for the effective theory describing ACR, including
the complete prefactors.}
\label{chiasym}
\end{figure}

Computing the impurity susceptibility is quite cumbersome,
since many graphs appear at intermediate stages of the
calculation, due to the fact that both pseudoparticles now
carry a spin index. First, we recall that the quantity which admits
a simple diagrammatics is in fact $\big<Q\big> \Cimp$ rather
than $\Cimp$, where  $\big<Q\big>$ is obtained from $\big<Q\big>
= e^{-\beta\lambda} \mcal{Z}$ and the above result for the
free energy, Eq.~(\ref{F_ACR}).
Again, we simply quote the collection of graphs
that need to be evaluated in Fig.~\ref{chiasym} (with the corresponding prefactor),
and refer the reader to Ref.~\onlinecite{fritz1} for more details.

\begin{thebibliography}{25}
\bibitem{epsfoot}
$\epsilon$ corresponds here
to the deviation of the parameter $r$ from its value at the lower-critical
or upper-critical dimension.
\bibitem{rmaxfoot}
Note that for $S=1/2$, the value
$r_\mr{max}=1/2$ corresponds rather to the collapse of SCR and SSC. The
critical point can then be captured by an expansion around SSC for
$r$ close to $1/2$ as shown in Ref.~\onlinecite{fritz1}, which is not
possible here since the collapse occurs at intermediate
coupling.
\bibitem{suscfoot}
We define the spin susceptibility with respect to
its $z$ component only due to spin rotation symmetry. For a
free spin $S$, we then have $\chi = S(S+1)/(3T)$.
\end{thebibliography}
\end{document}